# ESTemd: A Distributed Processing Framework for Environmental Monitoring based on Apache Kafka Streaming Engine

ADEYINKA AKANBI*

Centre for Sustainable SMART Cities

Central University of Technology, Free State, South Africa. 9300

aakanbi@cut.ac.za

Distributed networks and real-time systems are becoming the most important components for the new computer age – the Internet of Things (IoT), with huge data streams or data sets generated from sensors and data generated from existing legacy systems. The data generated offers the ability to measure, infer and understand environmental indicators, from delicate ecologies and natural resources to urban environments. This can be achieved through the analysis of the heterogeneous data sources (structured and unstructured). In this paper, we propose a distributed framework – Event STream Processing Engine for Environmental Monitoring Domain (ESTemd) for the application of stream processing on heterogeneous environmental data. Our work in this area is demonstrates the useful role big data techniques can play in an environmental decision support system, early warning and forecasting systems. The proposed framework addresses the challenges of data heterogeneity from heterogeneous systems and real-time processing of huge environmental datasets through a publish/subscribe method via a unified data pipeline with the application of Apache Kafka for real-time analytics.

## 1 INTRODUCTION

The recent surge of Internet of Things (IoT) has fueled the continues generation of real-time data from ubiquitous sensors and devices [14]. This sensorisation of the real world has lead to the wide adoption of real-time monitoring systems, from tsunami warning system [37], volcanic activity monitoring systems [38], smart cities [39], and recently in environmental monitoring and management. The application of real-time monitoring systems for environmental monitoring and management is essential because threats to public safety from environmental hazards such as drought, flooding and freakish weather events emerge regularly. This is shown by the 2019 splash flood in Johannesburg, South Africa [40], the 2019 Arkansas River floods in the US [41], and other international environmental events. Time-series data generated by real-time systems can become way too valuable at the time it is captured, processed and analysed in real-time to support valuable decision-making. However, the real-time analysis of data from environmental monitoring systems and other application areas is, however, challenged in several ways.

---

* Place the footnote text for the author (if applicable) here.

Firstly, the volume of data and sensory information currently produced by the heterogeneous sensors, devices and systems are huge, generated in an increasing speed with high data volume and represented in different underlying data models resulting in data heterogeneity [26]. Consequently, systems that produce the data and applications that are consumes the data are incompatible with each other [26]. In some cases, unstructured, semi-structured and structured data are produced by existing manual, legacy systems and IoT-based systems. Therefore, a mechanism is required for the seamless integration of data from producers and consumers for real-time analysis, irrespective of the data types. Secondly, the other challenge is managing the massive big data generated by these systems and performing real-time analytics on it for gaining meaningful insights without the need to commit the data to the database.

Recently, several technologies have emerged to address the challenges of real-time processing of high-volume streaming data using big data techniques. For example, complex event processing (CEP) systems and distributed stream processing (DSP) systems are typical examples (e.g. Apache Flink [1], Apache Kafka [23], Apache Storm [5], Google MapReduce [10], Hadoop [3,11], ESPER [12], SAP ESP [9], Microsoft StreamInsight [6], Siddhi [8], Apache Spark [7]). Stream analytics as a big data technology has shown great promise due to the implementation in a cloud environment and offers high throughput with low latency [34], [29]. However, in the environmental monitoring and management application domain, the combined use of data generated by legacy systems, relational database system dataset with streaming sensor data persists, and analysis of these heterogeneous datasets is crucial for accurate predictions and serves as inputs for numerical prediction models [28]. Therefore, in this study we describe Event STream Processing Engine for Environmental Monitoring Domain (ESTemd) framework – a framework that enables the integration of heterogeneous data sources with real-time analysis of the data using Apache Kafka stream processing engine for environmental monitoring domain. The presented framework solution is based on open-source Apache Kafka [23] in Confluent enterprise platform [25]. The data produced by heterogeneous devices are transformed and ingested using Apache Kafka Connect API and analysed by the Kafka stream processing engine based on a chosen numerical model for handling real-world events analysis. The proposed framework is extensible and compatible for use in other application domains.

The rest of the paper is organized as follows. The research background and existing related works are presented in section II. The current state-of-art and requirements analysis for the distributed streaming platform is discussed in section III. This is followed by the framework design, implementation, results and discussion with future work.

## 2 RELATED WORK AND BACKGROUND

Centralised computing systems have been around in technological computations for years [31]. In such systems, one central computer controls the peripherals and performs complex computations. Centralized computing systems require expensive hardware in order to process vast volumes of data and support multiple online users concurrently [20]. Under these circumstances, cloud and distributed computing systems arose to exploit parallel processing technology. Users can share computing resources, which can be virtualised and allocated dynamically in the cloud, typically characterised by scalable and elastic resources. With the advancement of IoT, the importance of stream processing systems increases as more and more modern applications impose tighter time constraints for real-time analysis of huge data generated [36, 38].

Stream processing (SP) is focused on analysing data streams from an event producer (e.g., sensors, devices, automated stations, relational database systems) using a data analytics platform (engine and infrastructure) to detect and extract meaningful insights, patterns and events in real-time without (the need of) committing this huge data stream to the database before processing. SP is important for real-time data analytics of continuous data streams from IoT sources [20],[21]. The huge volumes of data generated by IoT systems earned the title, 'Big Data'. These voluminous streams of sensor data are often characterised by the 5-Vs of Big Data – Volume, Variety, Value, Veracity and Velocity [22],[30],[31],[32]. The real value of such data is gained by new knowledge acquired by performing real-time data analytics using various data mining, machine learning or statistical methods [36].

There are several research approaches, which focuses on processing and analyzing streaming data using existing mentioned open-source big data analytics techniques [13],[14],[15],[16]. Malek et al. [14] introduced a platform design for real-time data stream monitoring and processing for the small-scale healthcare care to validate the viability of streaming processing in a critical sector. However, their work does not consider historical dataset in file format or heterogeneous data sources for processing. Rios [15] demonstrated the implementation of a system prototype for processing and analysing environmental sensor data in real-time, using Apache Storm for stream processing and Hadoop. This approach relies on homogenous data sources which are the sensor data alone, that makes it unreliable in the environmental monitoring domain due to use of dataset from legacy systems for modelling.

Motivated by these limitations and to overcome the challenges in the environmental monitoring application domain, in this research, we adopted open-source Apache Kafka implemented in a confluent enterprise platform [25] to eliminate complexities, with focus on heterogeneous data integration and stream processing of data using big data techniques. Apache Kafka makes it possible to process ingest heterogeneous datasets and data streams in real-time with the ability to react to notions and identify insights based on predictive algorithms [33].

Apache Kafka is an open-source distributed event streaming processing engine by Apache [33]. This streaming processing engine process sensor data streams in real-time to determine event patterns from incoming sensors' observation/readings and correlate the data with predefined/preset value threshold for prediction analysis. It is similar to an enterprise messaging system based on the ability to process sensor data streams in a fault-tolerant way as they occur in a producer-publish and consumer-subscribe fashion. Apache Kafka provides real-time processing of streaming data pipelines using persistent querying systems (KSQL) without the need to commit the data stream to the database like conventional systems [24]. This provides a huge benefit in IoT-enabled environmental monitoring systems for real-time monitoring of complex environmental phenomenon. Kafka provides a real-time publish-subscribe solution that overcomes the challenges of consuming the real-time and batch dataset volumes that may grow in the order of magnitude to be larger than the real data.

## 3 REQUIREMENT ANALYSIS AND STATE OF THE ART

Previously, in an effort for effective analysis of big data, batch processing emerged as a research area, saddled with the goal of analysing huge dataset in batches – collected over a period of time – to extract meaningful insights, patterns and events [17]. This is achieved through the processing of raw data coming from diverse data sources and processed through a batch processing engine based on a predefined model or logic to identify likely events or future scenarios. Batch processing technologies are based on the extract-transform-load (ETL) principle, that causes latency due to low processing and analysis speed. Typical examples are Hadoop [17] and Google MapReduce [18] – used to process a large volume of data sources in a batch-oriented way.

However, Big data analytics has seen a radical shift from near-real-time using batch processing to real-time stream processing [36], which improves the processing capabilities and reduce the amount of time dramatically [33]. Although batch processing tools are such as Hadoop are efficient, there is an increasing demand for real-time processing and analysis of big data streams [17]. Realtime stream processing technologies, such as Storm [5], Spark [2], Kafka [23], have been developed to process data while in motion in order to get as fast as possible valuable insights from it.

A typical state-of-art big data analytics portfolio consists of four layers stack [19], namely: (1) Data sources, (2) Data Ingestion layer, (3) Data analytics layer, and (4) Storage layer as depicted in Figure 1 below:

- Data Sources: This layer consists of the producers or the devices such as sensors and input systems generating the data or the data streams. Standard ingestion layers also have API that facilitates data communication for seamless data integration and perform simple data transformation. The layer also buffers the input data and make it available for further processing, best described as a producer-consumer;
- Data Ingestion Layer: The ingestion layer is a critical layer in any big data platform or infrastructure. This layer is saddled with the responsibility of ingestion heterogeneous data sources using compatible APIs for processing by the stream analytics layer.
- Data Analytics Layer: This layer is responsible for consuming the streaming data or input data from the data ingestion layer for processing and analysis. Modern Big Data platform analytics layer incorporates numerical models and algorithm.
- Storage Layer: Unlike the other layers, the storage layer is responsible for persistent storage of the aggregated input and output data streams depending on the framework or for archiving the streams for further usage in a fault-tolerant way.

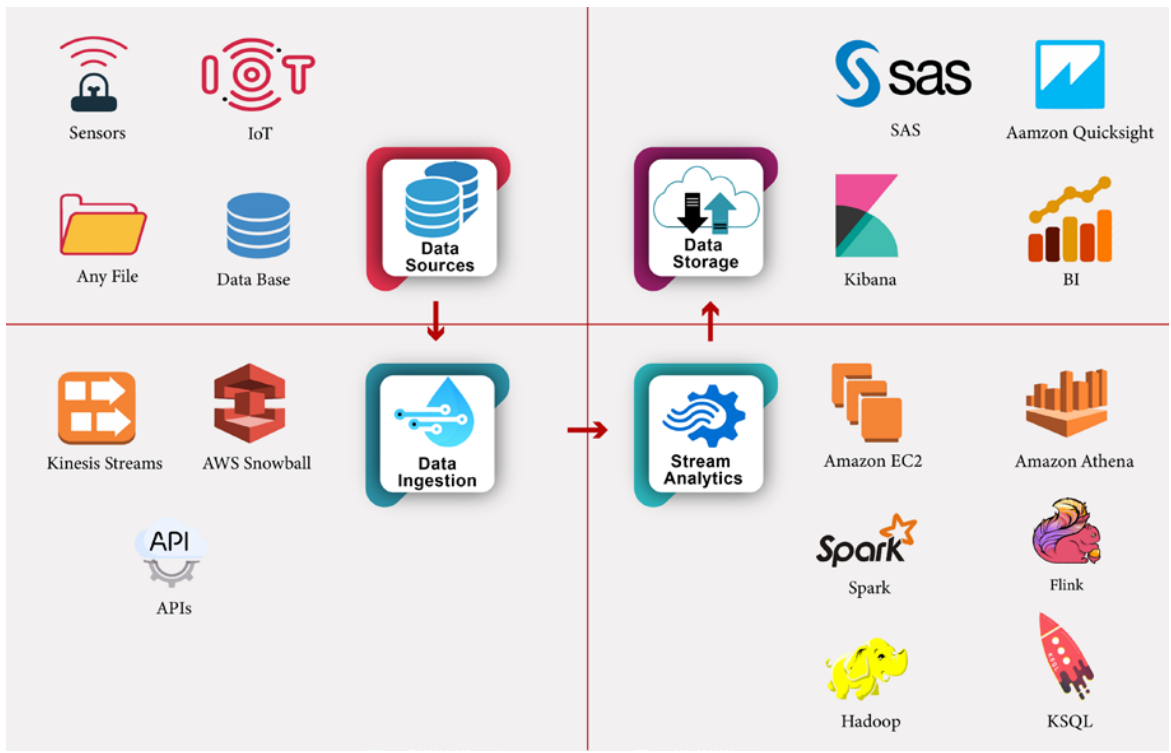

Figure 1: Generic Big Data analytics portfolio four-layers stack [42].

## 4 ESTEMD FRAMEWORK DESIGN

### 4.1 ESTemd High-level Architecture

In this section, we propose a distributed stream processing framework called Event STream Processing Engine for Environmental Monitoring Domain (ESTemd), that integrates a collection of suitable technologies from the existing state of the art work. We describe the framework and how it handles the challenges of real-time analysis of data from environmental monitoring and management systems discussed previously.

The framework addresses the complex task of data integration from distributed heterogeneous sources and stream processing by using open-source Apache Kafka in Confluent platform for large scale data processing and real-time intelligence. In this way, a framework that is reusable, scalable, that employs a multi-tenant approach and compatible with current big data technologies are developed. The framework design satisfies the requirement for efficient data processing for IoT applications and supports the extraction of insights from a stream of incoming sensors observation and unstructured data sources. Figure 2 illustrates the ESTemd framework stack for heterogeneous data integration and big data analysis

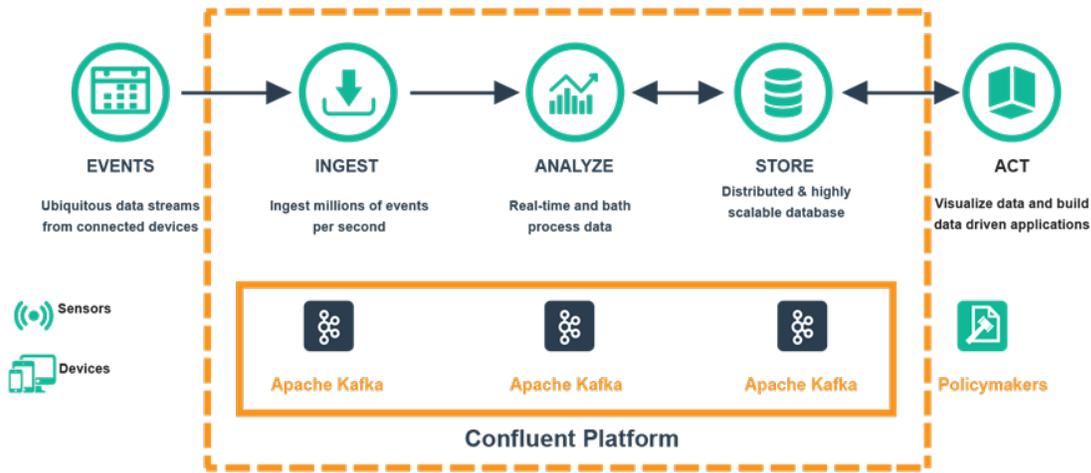
Figure 2: High-level architectural stack of the ESTemd framework.

### 4.2 ESTemd Framework Modules

*4.2.1 Data Ingestion Layer.*

This layer is the open platform used by sensors, devices, automated weather stations, legacy weather stations to measure and record meteorological parameters. The gathered environmental datasets are made available through a central endpoint or communication link for access by the next layer. Specifically, the layers collect data from different data sources (structured and unstructured) and adopt varieties of technology or tool as a means for effective data gathering. This layer must be a highly scalable using a publish-subscribe event bus which ensures that data streams are captured with minimal loss. In this research, Confluent enterprise platform provides the use of Apache Kafka source connectors acting as a broker to buffer the incoming data streams from the producers and also helps to achieve better fault tolerance and load balancing in the eventuality of component failure [29]. The data are channelled from the producers to the next layer (brokers) of the framework via respective Kafka topics in the cluster – ready to be queried or utilised by the streaming engine within the Confluent environment. The data stream from the producers is responsible for feeding the system.

*4.2.2 Data Broker Layer.*

This layer provides the necessary technological solution or means to access and transform the heterogeneous data gathered by the producers using standardised APIs with wide compatibility range. The data broker layer is sub-divided into two types: Apache Kafka Connect Source API and the Apache Kafka Connect Sink API – collectively called the Kafka streaming APIs [27]. The Kafka Connect Source API receives the data from the previous layer and makes real-time light-weight modifications to the raw messages to make the data compatible with the unified data pipeline for seamless integration with the Apache streaming processing engine in the Confluent ecosystem. Apache Kafka Connect Source API acting as a broker will buffer the incoming data streams from the producers. Kafka Source API is available for MQTT, Kafka Connect RabbitMQ, Kafka Connect JDBC, Kafka Connect CDC Microsoft, RabbitMQ, HDFS, HTTP, MongoDB, Neo4j, Cassandra.

The Kafka Connect Sink API work exactly like the previously discussed Kafka Connect Sink API. Kafka Stream APIs forms the data broker layer and works on either side of the Kafka streaming processing engine, as illustrated in Figure 3 below. The output from the stream processing engine is made available to other clusters using Kafka Connect sink connector API. The Sink Connect API provides the communication connection protocol for the transformation of the data from the data pipeline compatible format for the storage layer. The Kafka Sink Connector acts as a buffer to output from the streaming engine, using the Kafka Connect Single Message Transform to make lightweight modifications to the Kafka messages before writing the outputs to secondary indexes for an offline longer time series analysis or immediate visual analysis using AKKA [35], Microsoft Bi [34] for further insights.

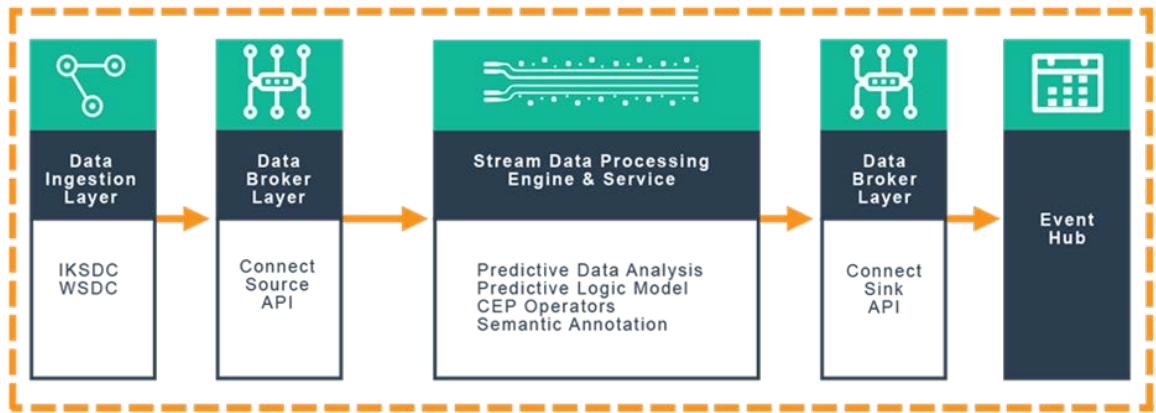

Figure 3: A layered view of the ESTemd framework.

### 4.2.3 Stream Analytics Layer.

The streaming analytics layer normally incorporates the stream processing engine responsible for real-time analysis of input data streams; and model management responsible for maintaining the numerical models used for persistent querying of the data streams. The stream processing engine makes use of CEP operators to identify meaningful patterns, relationships and gain predictive insights from streams of unbounded dataset using operators such as Filter (), Map (), FlatMap (), Aggregation (), Projection (), Negation () based on the numerical model [21]. Using the Confluent platform, streams of data categorised based on similar attributes are streamed to the Kafka topics in the cluster broker to be processed by the Kafka-SQL (KSQL) node to detect events in the time-series data streams. KSQL is a SQL-like querying language for Apache Kafka and designed to meet the main requirement of persistent querying of the input streams [24].

### 4.2.4 Storage Layer.

This layer is responsible for storing the output of the real-time data analysis. In most case, output data streams are committed to appropriate Kafka topics and saved to multiple elasticsearch indexes or data storage such as distributed file storage (DFS), Hadoop Distributed Filesystem (HDFS), structured data sources, NoSQL, Cassandra, MongoDB, Neo4js, HBase etc.

## 5 FRAMEWORK IMPLEMENTATION

The implementation of the ESTemd framework in the environmental management and monitoring domain will help to aggregate heterogeneous data, process and analyse any data streams or dataset in real-time using big data techniques. One of the implemented use cases is a real-time meteorological monitoring system for the determination of effective precipitation. Effective Precipitation (EP) is the amount of precipitation that is added and stored in the soil. Sensors (producers) measures the precipitation values; with readings data streamed to designated Kafka topic are ingested and analysed by the stream processing engine based on the EP model algorithm using the EP equation (1) below.

$$Effective\ Precipitation\ (mm) = (RAIN - 5)\ x\ 0.75 \qquad (1)$$

### 5.1 Data

The data used in this study was collected from a calibrated weather station and a deployed wireless sensor network (WSN) during the month of April 2020 at Bloemfontein, Free State Province, South Africa. The raw precipitation data from the producers (weather stations & sensors) was collected in a 5-min interval. The data from the weather station was in CSV format, and the data from the WSN in JSON format; both indicating the heterogeneity of the data from the producers. More details on the experimental setup and data collection methods are provided in [42].

### 5.2 Methods

The framework is implemented as an infrastructure containing one cluster running on a local machine with a quad-core Intel CPU and 8GB RAM hosts in Confluent platform with Zookeeper, an instance of Kafka broker, an active controller, and Kafka client hosting the Kafka streaming engine API and the KSQL for persistent querying of the streams in real-time. The sensor rain measurement dataset in JSON and CSV format are ingested by the system to the Kafka RAIN topic using the JDBC Connect API (Figure 4).

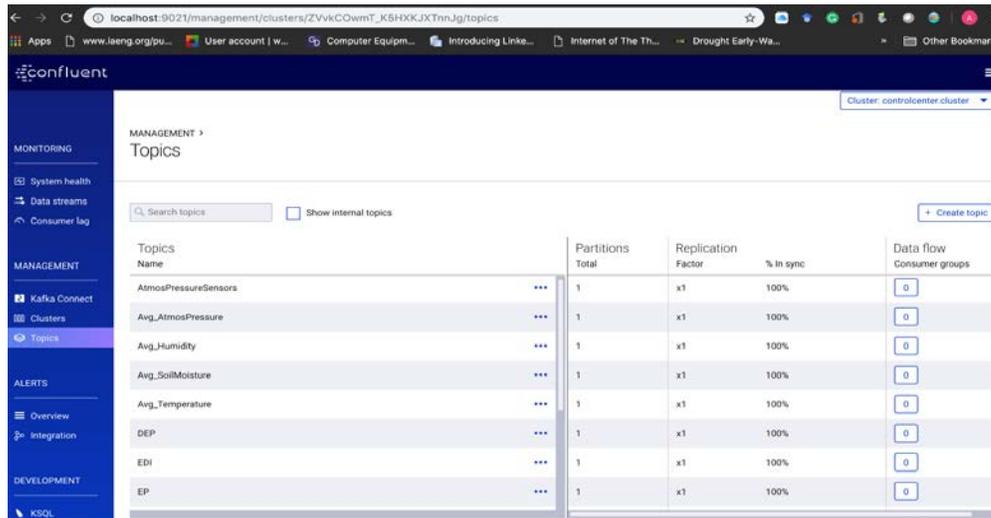

Figure 4: Kafka topics created in the broker.

In this study, a unique topic (Rain) is created to cater for the precipitation readings from the producers (sensors). Further manipulation of the Kafka topic messages using CEP operators based on the EP model formula will yield new topic (Precipitation) to store the output of the processed messages.

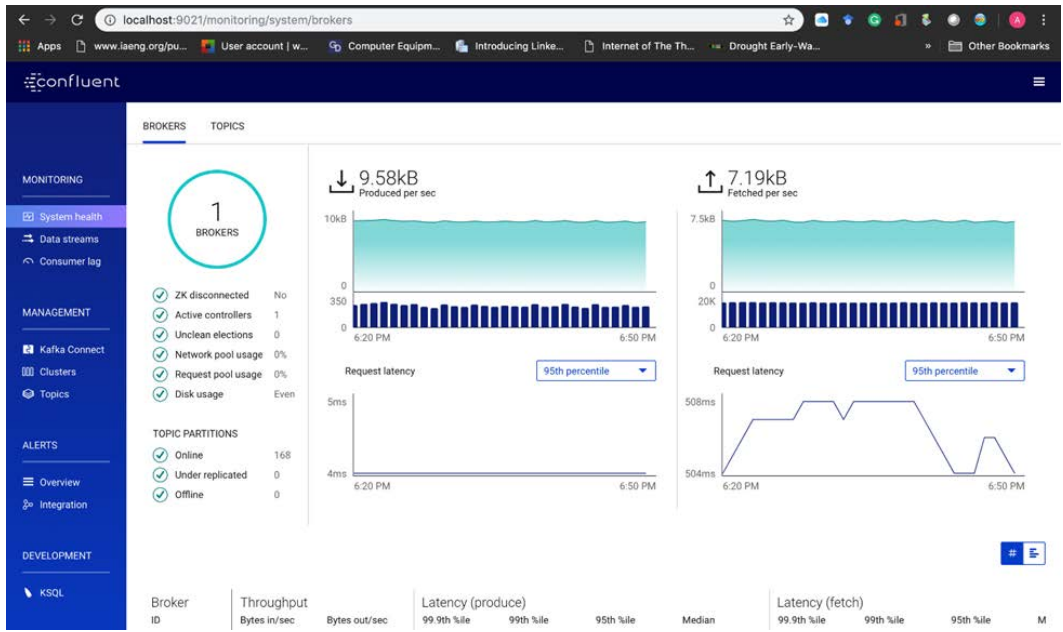

Figure 5: The visualisation output from the Confluent streaming platform.

The data streams generated by the sensors (producers) are passed on to the Kafka topic in the Kafka broker for stream processing. The Kafka broker runs operators and user-defined functions inside the JVM. Computational model processes are performed on the data streams using KSQL to generate a result that will be committed to the output topics in the broker. Each record or message from a producer is typically represented as a key-value pair, and the streams of record are processed in real-time with the smallest amount of latency through the help KSQL. Moreover, the cluster is monitored and managed through the Confluent streaming platform and accessible through port 9021 (Figure 5). The entire input data streams are queried through the KSQL (Figure 6) based on the algorithm of the EP equation 1. The output of the processed stream is store in the storage layer.

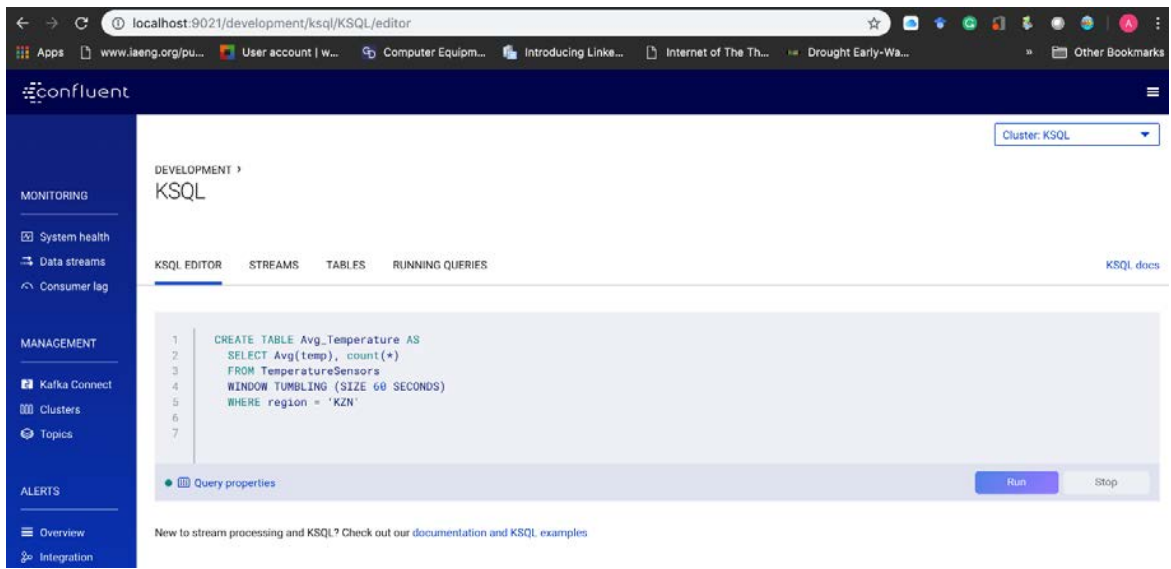

Figure 6: Querying the streams using KSQL.

## 6 RESULT AND DISCUSSION

The output of the processed input streams is available through the designated output topic and saved in the event hub for further availability or visualisation of the data through appropriate plugins. In this case study, the experiment was performed in an area where the rainy season is evenly distributed throughout the year. At the moment of implementation, the output value of the persistent analysis and processing of the input streams based on the numerical computational model indicates an EP average value of 211.8 (Figure 7). The output results are subjected to interpretation by the domain expert.

```
ksql> SELECT Value FROM EP;
2020-04-02 | 17:05:07 | 210.00 | EP
2020-04-02 | 17:05:07 | 211.80 | EP
2020-04-02 | 17:05:07 | 211.81 | EP
2020-04-02 | 17:05:07 | 211.80 | EP
2020-04-02 | 17:05:07 | 211.81 | EP
Limit Reached
Query terminated
```

Figure 7: The output stream from the persistent querying of the input streams through the KSQL editor.

The monitoring of an environmental phenomenon with time series of data and making them available for studying are the key roles of environmental monitoring systems. In this paper, we first motivated the need and discussed the challenges of applying big data techniques for real-time analysis of heterogeneous environmental datasets. We propose ESTemd, a distributed stream processing framework for real-time analysis of heterogeneous environmental monitoring data using Apache Kafka in Confluent environment. The novelty of our approach is to provide a decentralised system that is based on several technological solutions to solve the challenges of data heterogeneity by hiding the complexities of various data types through the use of a unified data. The proposed framework is extensible and applicable to other research areas, thanks to the loosely coupled modular architecture. The ESTemd framework is applicable for real-time monitoring systems or for decision support for gaining insights from huge datasets. Our future work will focus on the framework implementation using a complex numerical model on diverse data sources in an enterprise environment. Additionally, we are exploring the semantic representation of the large-scale time series data using appropriate domain ontology.